\documentclass{jfm}
\usepackage{amsmath,amsfonts,amssymb,bm,upmath,graphicx}
\newcommand{\ud}{\mathrm{d}}
\newcommand{\ue}{\mathrm{e}}
\newcommand{\ui}{\mathrm{i}}

\title{Renormalized transport of inertial particles in surface flows}
\author[M. Martins Afonso, A. Mazzino and P. Olla]{M\ls A\ls R\ls C\ls O\ns M\ls A\ls R\ls T\ls I\ls N\ls S\ns A\ls F\ls O\ls N\ls S\ls O$^1$,
 A\ls N\ls D\ls R\ls E\ls A\ns M\ls A\ls Z\ls Z\ls I\ls N\ls O$^2$
 \and P\ls I\ls E\ls R\ls O\ns O\ls L\ls L\ls A$^3$}
\affiliation{$^1$Department~of~Mechanical~Engineering, Johns~Hopkins~University, Baltimore,~MD~21218,~USA\\[\affilskip]
 $^2$Department~of~Physics~-~University~of~Genova, and CNISM \& INFN~-~Genova~Section, via~Dodecaneso~33, 16146~Genova, Italy\\[\affilskip]
 $^3$ISAC-CNR \& INFN~-~Cagliari~Section, 09042~Monserrato~(CA), Italy}
\date{\today}

\usepackage{natbib}
\allowdisplaybreaks[1]

\begin{document}

 \maketitle

 \begin{abstract}
  Surface transport of inertial particles is investigated by
  means of the perturbative approach, introduced by Maxey
  ({\it J.\ Fluid Mech.}\ {\bf 174}, 441 (1987)),
  which is valid in the case the deflections induced on
  the particle trajectories by the fluid flow can be
  considered small. We consider a class of compressible
  random velocity fields, in which the effect of recirculations is
  modelled by an oscillatory component in the Eulerian time
  correlation profile.
  The main issue we address here is whether fluid velocity fluctuations,
  in particular the effect of recirculation,
  may produce nontrivial corrections to the streaming particle velocity.
  Our result is that a small (large)
  degree of recirculation is associated with a decrease (increase)
  of streaming with respect to a quiescent fluid.
  The presence of this effect is confirmed numerically, away from the perturbative limit.
  Our approach also allows us to calculate the explicit expression for the eddy diffusivity,
  and to compare the efficiency of diffusive and ballistic transport.
 \end{abstract}

 \section{Introduction}

 Particle transport in laminar/turbulent flows is a problem of
 major importance in a variety of domains ranging from astrophysics
 to geophysics. For neutral (i.e.\ having the same density
 of the surrounding fluid) particles in incompressible flows the main
 quantity of interest is typically the rate at which the flow
 transports the scalar, e.g.\ a pollutant. For times large compared
 to those characteristic of the flow field, transport is diffusive
 and is characterized by effective diffusivities (the so
 called eddy diffusivities) which incorporate all the nontrivial
 effects played by the small-scale velocity on the asymptotic
 large-scale transport \cite[][]{BCVV95,AM97}. Eddy diffusivities are expected on the
 basis of central limit arguments and can be calculated by means of
 asymptotic methods \cite[][]{BO78}. There are however situations where some,
 or all, of the hypotheses of the central-limit theorem break down
 with the result that in the asymptotic limit particles do not
 perform a Brownian motion and anomalous diffusion is observed
 \cite[][]{CMMV99,ACMV00}.\\
 For compressible flows, large-scale transport is still controlled
 by eddy diffusivities but now transport rates are enhanced or
 depleted depending on the detailed structure of the velocity
 field \cite[][]{VA97}.\\
 For particles much heavier than the surrounding fluid,
 large-scale transport is still controlled
 by eddy diffusivities, which have been calculated in \cite{PS05}
 exploiting asymptotic methods in the limit of small inertia.\\
 Unlike what happens for neutral particles, an external force field
 can dramatically change transport properties. This is for instance
 the case when gravity is explicitly taken into account, with the
 result that a constant falling/ascending velocity sets in, thus
 dominating the particle long-time transport
 \cite[][]{MC86,M87a,M87b,M90,ACHL02,FK02,RMP04,MFS07,MA08}.\\
 Our main aim here is to focus on transport of inertial
 particles in compressible media. Although our results are rather general
 (i.e.\ they apply to both one, two and three dimensions, and for a
 variety of external forces), the present study is physically
 motivated by the transport of floaters on a surface flow (e.g.\ the
 ocean surface) in the presence of strong surface winds. As we
 will see, the resulting dynamical equations for the particles
 moving on the (horizontal) surface under the action of strong
 (constant) winds are formally identical to those of heavy
 particles moving along the vertical under the action of gravity.
 A constant drift is thus expected, analogous to the settling
 velocity of a particle in suspension, in the presence of gravity. \\
 The specific problem we aim at investigating can be summarized
 as follows. The drift velocity $\bm{v}$ of a floater on a still water
 surface will be in general a complicated function of the wind
 velocity, $\bm{U}$, and of other parameters like e.g.\ the
 floater structure and the surface roughness. A similar complication is
 expected in the way in which the drift velocity adapts to the wind
 and surface current variations. A rather minimal model could be
 obtained assuming a simple relaxation dynamics for the drift velocity:
 \begin{equation} \label{model}
  \dot{\bm{v}}=[\bm{\mathcal{V}}-(\bm{v}-\bm{u})]\gamma,
 \end{equation}
 where $\bm{u}$ is the water surface velocity,
 $\gamma=\gamma(\bm{U}-\bm{u},\bm{v}-\bm{u})$
 is the relaxation rate, and $\bm{\mathcal{V}}$ is the
 terminal drift velocity (with $\bm{\mathcal{V}}=\bm{\mathcal{V}}(\bm{U}-\bm{u})$).
 Since we expect $\mathcal{V},u\ll U$, we may approximate
 $\bm{U}-\bm{u}\simeq\bm{U}$
 in the arguments of both $\bm{\mathcal{V}}$ and $\gamma$, and, as a rough
 approximation, we may consider a linear relaxation dynamics:
 $\gamma(\bm{U},\bm{v}-\bm{u})\simeq\gamma(\bm{U},0)$.
 Under these hypotheses, (\ref{model}) becomes formally identical
 to the one for a small heavy particle in a viscous fluid,
 in the presence of a gravitational acceleration $\gamma\bm{\mathcal{V}}$.
 In this case, $\gamma^{-1}$ would be the Stokes time \cite[][]{MR83,M97}.\\
 Now, spatio-temporal variations in the surface water velocity
 will lead to difficulties in the determination of a mean drift
 velocity, due to preferential concentration effects \cite[][]{M87b}.
 In other words, the mean drift may not necessarily be equal
 to what would be obtained by calculating a spatio-temporal average.
 In general, the interplay between currents and particle trajectories might
 lead to the result that either an enhanced or a reduced drift
 (with respect to the one in still fluid) might appear.\\
 Here, we will assess the above possibility by means of an
 analytical (perturbative) approach in the same spirit of \cite{M87b}. We
 will be able to identify an important dynamical feature of the
 carrying flow (the way through which it decorrelates in time)
 responsible of the different behaviour of the resulting drift with
 respect to the corresponding value in still fluid.

 The paper is organized as follows. In \S~\ref{sec:ge} the basic equations
 governing the time evolution of inertial particles in a prescribed flow are
 given, together with the perturbative expansion for strong sweeping or gravity.
 In \S~\ref{sec:cf} we focus on compressible flows and compute the leading
 correction to terminal velocity. The same quantity is calculated in \S~\ref{sec:igf}
 for incompressible Gaussian flows. In \S~\ref{sec:pd} we analyse the phenomenon
 of particle diffusivity and, at the leading order, we study the diffusion coefficient;
 we also perform a quantitative comparison of the drift and diffusion displacements.
 Conclusions follow in \S~\ref{sec:c}. The appendices are devoted (\S~\ref{sec:a1})
 to displaying some additional analytical results, and (\S~\ref{sec:a2}), to provide
 some technical details of the calculation.

 \section{General equations} \label{sec:ge}

 In the hypotheses of (\ref{model}), the motion of a floater dragged by the wind,
 on a water surface with velocity field $\bm{u}(\bm{x},t)$, will be described by:
 \begin{equation} \label{eq1}
  \left\{\begin{array}{l}
   \dot{\bm{x}}(t)=\bm{v}(t)\\
   \dot{\bm{v}}(t)=\gamma\,[\bm{\mathcal{V}}+\bm{u}(\bm{x}(t),t)-\bm{v}(t)]\;,
  \end{array}\right.
 \end{equation}
 with the quantities $\bm{\mathcal{V}}$ and $\gamma$ assumed as constants.
 We shall denote the direction of the vector $\bm{\mathcal{V}}$ as ``sweeping'',
 and we align the axes such that it corresponds to the positive $x_d$ component. To make
 contact with the dynamics of a heavy particle in a viscous fluid,
 we maintain $d$, that is the number of dimensions, arbitrary.\\
 Given (Eulerian) characteristic length and time scales
 $L$ and $T$ and amplitude $\sigma_u$  for the velocity field $\bm{u}$,
 we introduce the dimensionless Stokes, Kubo and
 Froude numbers $S$, $K$ and $F$, defined as
 \begin{equation} \label{eq1.1}
  S=\frac{1}{\gamma T},\qquad K=\frac{\sigma_u T}{L}\quad\textrm{and}\quad F=\frac{\sigma_u}{\sqrt{\gamma\mathcal{V}L}}\;.
 \end{equation}
 Notice that such a definition of $F$ is consistent with the
 ``identification'' $\bm{g}=\gamma\bm{\mathcal{V}}$, if gravity is the
 ``sweeping'' force under consideration (rather than wind).

 From now on, we adimensionalize times with $\gamma^{-1}$ and velocities with $\sigma_u$,
 and denote the new adimensional variables with the same letters as before.
 In dimensionless form, the characteristic scales $L$ and $T$ of the velocity field,
 and the bare terminal velocity $\mathcal{V}$, will read:
 \begin{equation} \label{eq1.2}
  L=S^{-1}K^{-1},\qquad T=S^{-1}\quad\textrm{and}\quad\mathcal{V}=SKF^{-2}\;.
 \end{equation}
 The Kubo number $K$ is basically the ratio of the life time and rotation
 time of a vortex, with $K\to0$ corresponding to an uncorrelated
 regime, $K\to\infty$ to a frozen-like regime, and real turbulence being
 realized by $K=O(1)$.\\
 We confine ourselves to situations in which the particle
 trajectory, after some transient regime depending on the initial conditions,
 shows small deviations from the ``sweeping'' line, thus allowing us
 to deal with a quasi-1D problem as a zeroth-order approximation \cite[][]{M87b}.
 This happens when the effect of streaming (or gravity) is much stronger than the
 deflections due to the external flow.\\
 We thus isolate in the solution $\bm{v}=\bm{v}(t)$ of (\ref{eq1})
 a term associated with the deviation from the behaviour in still fluid:
 \begin{equation} \label{vtilde}
  \tilde{\bm{v}}(t)=\!\int_0^t\ud t'\,\ue^{t'-t}\bm{u}(\bm{x}^{\scriptscriptstyle(0)}(t')+\tilde{\bm{x}}(t'),t')\;,
 \end{equation}
 where
 \begin{equation} \label{xbar}
  \bm{x}^{\scriptscriptstyle(0)}(t)=\bm{x}(0)+\bm{\mathcal{V}}t+[\bm{v}(0)-\bm{\mathcal{V}}](1-\ue^{-t})
 \end{equation}
 accounts for the unrenormalized sweep, and $\ud\tilde{\bm{x}}/\ud t=\tilde{\bm{v}}$, so that:
 \begin{equation} \label{xtilde}
  \tilde{\bm{x}}(t)=\!\int_0^t\ud t'\,\psi(t-t')\bm{u}(\bm{x}^{\scriptscriptstyle(0)}(t')+\tilde{\bm{x}}(t'),t'),\qquad\psi(t)=1-\ue^{-t}\;.
 \end{equation}
 A perturbative solution of (\ref{xtilde}) rests on the
 smallness of $\tilde{\bm{x}}(t)$ in the argument of $\bm{u}$. More
 precisely, it is necessary that $\tilde{x}(\tau_{\mathrm{p}})\ll
 L$, with $\tau_{\mathrm{p}}$ the correlation time for the fluid
 velocity $\bm{u}(\bm{x}(t),t)$ sampled by the particles. We can estimate
 \begin{equation} \label{T_sweep}
  \tau_{\mathrm{p}}\sim\min(T,L/\sigma_u,\mathcal{T}_{\mathrm{sw}}),\qquad\mathcal{T}_{\mathrm{sw}}\equiv L/\mathcal{V}=S^{-2}K^{-2}F^2\;,
 \end{equation}
 with $\mathcal{T}_{\mathrm{sw}}$ giving the contribution from
 sweep to decorrelation. In the absence of sweep effects, we know
 that $\tilde{x}(\tau_{\mathrm{p}})/L$ will be small provided
 either $K\ll1$, or $K\gtrsim1$ with $SK^{-2}\gg1$
 \cite[][]{WM03,OV07}. In both regimes, indeed, the inertial particles
 will see $\bm{u}$ as a Kraichnan field \cite[][]{K68,K94}. Sweeping acts to
 reduce the correlation time. A sufficient condition for small
 $\tilde{x}(\tau_{\mathrm{p}})/L$ turns out to be, in this case:
 \begin{equation} \label{condition}
  SKF^{-2}\gg1\;,
 \end{equation}
 that is the strong-sweep condition $\mathcal{V}\gg\sigma_u$ (no
 condition is put on $\gamma$). Notice that, for sufficiently
 large $S$, this condition and the one for a Kraichnan regime $K\ll1$ overlap.
 We assume $\bm{u}$ as a homogeneous, isotropic, stationary, zero-mean random flow
 and we denote with $\langle\cdot\rangle$ the ensemble average over its
 realizations. We are interested in finding the steady-state average particle velocity,
 which corresponds, in (\ref{vtilde}), to averaging over $\bm{u}$ and taking $t\to+\infty$.\\
 Taylor expanding in $\tilde{x}$ the right-hand side (RHS) of (\ref{vtilde}), and using
 (\ref{xtilde}) recursively, allows us to calculate the correction to sweep.
 We clearly obtain $\langle\tilde{\bm{v}}^{\scriptscriptstyle(1)}(t)\rangle=0$.
 Then, we have
 \begin{equation} \label{v1}
  \langle\tilde{v}_i^{\scriptscriptstyle(2)}(t)\rangle=\!\int_0^t\ud t'\,\psi(t-t')\langle\mathcal{U}_j(t')\partial_j\mathcal{U}_i(t)\rangle\;,
 \end{equation}
 where
 \begin{equation} \label{mathcal}
  \bm{\mathcal{U}}(t)=\bm{u}(\bm{x}^{\scriptscriptstyle(0)}(t),t)
 \end{equation}
 (i.e.\ the unperturbed flow is ``sampled'' on the \emph{fixed} ``sweeping'' line).
 As it is well known, the lowest order correction to $\mathcal{V}$ vanishes
 if $\bm{u}$ is incompressible. In this case, in order to obtain non-zero corrections
 to the falling velocity, it is necessary to
 go to higher orders in the perturbative expansion of (\ref{vtilde})--(\ref{xtilde})
 \cite[][]{M87b}. Namely, in the incompressible case, we have:
 \begin{equation} \label{nuova}
  \langle\tilde{v}_i^{\scriptscriptstyle(3)}(t)\rangle=\!\int_0^t\ud t'\,\psi(t-t')\!\int_0^{t'}\ud t''\,[\psi(t'-t'')-\psi(t-t'')]\langle\mathcal{U}''_k\mathcal{U}'_j\partial_j\partial_k\mathcal{U}_i\rangle\;.
 \end{equation}
 The latter quantity is zero if $\bm{u}$ is a Gaussian field.
 Thus, for incompressible Gaussian flows, one has to compute the next order:
 \begin{eqnarray} \label{v3}
  \langle\tilde{v}_i^{\scriptscriptstyle(4)}(t)\rangle&=&\!\int_0^t\ud t'\int_0^{t'}\ud t''\,\psi(t-t')\psi(t'-t'')\times\\
  &&\times\left[\!\int_0^t\ud t'''\,\psi(t-t''')-\!\int_0^{t''}\ud t'''\psi(t''-t''')\right]\partial_j\partial_l\langle\mathcal{U}_i\mathcal{U}''_k\rangle\partial'_k\langle\mathcal{U}'_j\mathcal{U}'''_l\rangle\;.\nonumber
 \end{eqnarray}
 In \S~\ref{sec:cf} we provide an example of compressible flow,
 for which (\ref{v1}) applies. Due to the difficulty of dealing analytically with
 non-Gaussian flows, we will not provide applications of (\ref{nuova}).
 On the contrary, in \S~\ref{sec:igf}, we will focus on an incompressible Gaussian flow,
 for which (\ref{v3}) is the leading correction to the falling velocity.

 \section{Compressible flows} \label{sec:cf}

 Let us focus on the compressible case and compute the average
 in (\ref{v1}). Clearly we only need to compute the component with index
 $i=d$ (in our convention it is
 assumed positive if pointing along the mean flow).\\
 Let us first analyse $d>1$.
 Introducing the well-known compressibility degree,
 $\mathcal{P}\equiv\langle(\partial_ju_j)^2\rangle/\langle(\partial_ku_l)(\partial_ku_l)\rangle\in[0,1]$,
 the expression for the Eulerian pair correlation tensor $R_{ij}(\bm{x},t)\equiv\langle u_i(\bm{x},t)u_j(\bm{0},0)\rangle$
 can be deduced from
 \begin{eqnarray} \label{uij}
  R_{ij}(\bm{x},t)&=&\left[(1-d\mathcal{P})\partial_i\partial_j-(1-\mathcal{P})\delta_{ij}\partial^2\right]\mathcal{R}(x,t)
 \end{eqnarray}
 by imposing the form of the scalar $\mathcal{R}(x,t)$. Let us assume a Gaussian behaviour both in
 space and in time, with temporal oscillations described by
 $\omega$ (adimensionalized with $\gamma$ for the sake of consistency):
 \begin{equation} \label{cxt}
  \mathcal{R}(x,t)=\frac{1}{d(d-1)S^2K^2}\cos(\omega t)\ue^{-S^2t^2/2}\ue^{-S^2K^2x^2/2}\;.
 \end{equation}
 Substituting into (\ref{uij}), we obtain
 \begin{eqnarray} \label{uijxt}
  R_{ij}(\bm{x},t)&=&\frac{1}{d(d-1)}\cos(\omega t)\ue^{-S^2t^2/2}\ue^{-S^2K^2x^2/2}\times\\
  &&\times\left\{(1-d\mathcal{P})S^2K^2x_ix_j+\delta_{ij}\left[(d-1)-(1-\mathcal{P})S^2K^2x^2\right]\right\}\nonumber
 \end{eqnarray}
 and thus
 \begin{equation} \label{duijxt}
  \partial_jR_{ij}(\bm{x},t)=\frac{\mathcal{P}S^2K^2}{d}\cos(\omega t)\ue^{-S^2t^2/2}\ue^{-S^2K^2x^2/2}\left[S^2K^2x^2-(d+2)\right]x_i\;.
 \end{equation}
 The case $d=1$ automatically implies $\mathcal{P}=1$, and
 expressions (\ref{uij}) through (\ref{uijxt}) are ill posed. However,
 if one neglects them and considers expression (\ref{duijxt})
 directly, everything is consistent and also the 1D
 case can be investigated by means of the same formalism. Our
 choice of obtaining the latter equation passing through the
 former three for $d>1$ is simply dictated by the simple and important meaning
 that the compressibility degree plays: the lower bound $\mathcal{P}=0$ denotes
 incompressible flows and the upper bound $\mathcal{P}=1$
 perfectly compressible (potential) ones.\\
 It is interesting to study under what conditions the tensor (\ref{uijxt})
 is positive definite (for $d>1$). One can notice that, as in our calculations
 it always appears under some integral, it is sufficient to study
 the separations $x<L$, because the Gaussian factor makes the
 contribution from larger distances negligible. Thus, using
 (\ref{eq1.2}), $S^2K^2x^2|_{x<L}<1$, and therefore the quantity
 in square braces beside $\delta_{ij}$ is always positive in this case.
 Then, the investigation of the tensorial part suggests that
 $R_{ij}$ would surely be positive definite (for the only relevant
 interval, $x<L$) when $\mathcal{P}<d^{-1}$,
 if the time-oscillating factor were absent. This tells us that,
 at low compressibility, the introduction of the temporal oscillations
 corresponds to the appearance of negatively-correlated areas,
 and tuning the parameter $\omega$ one can change the relevance of the recirculating flow.

 Under our assumptions, performing the simple change of variables
 $\tau\equiv t-t'$, at very large times $\forall d$ expression
 (\ref{v1}) for the ``sweeping'' component simplifies to
 \begin{equation} \label{vd}
  V\equiv\lim_{t\to\infty}\langle\tilde{v}_d^{\scriptscriptstyle(2)}(t)\rangle=\frac{\mathcal{P}S^3K^3}{dF^2}\!\int_0^{\infty}\ud\tau\,\tau\left[\frac{S^4K^4}{F^4}\tau^2-(d+2)\right]\cos(\omega\tau)\psi(\tau)\ue^{-\Gamma^2\tau^2/2}\;,
 \end{equation}
 where
 \begin{equation} \label{Gamma}
  \Gamma=S\sqrt{1+\Delta^2}\approx\tau_{\mathrm{p}}^{-1}\quad\textrm{and}\quad\Delta=SK^2F^{-2}
 \end{equation}
 are respectively the decorrelation rate, and the ratio $T/\mathcal{T}_{\mathrm{sw}}$.

 The last integral in (\ref{vd}) can be carried out analytically and the result
 can be expressed in terms of error functions, but for the sake of
 simplicity it is not reported here (see (\ref{eqap}) in appendix \ref{sec:a1}).
 Out of the six parameters in play, we notice that
 $V$ is exactly linear in $\mathcal{P}$, and coherently
 vanishes in the incompressible case, thus we can fix $\mathcal{P}=1$
 for the rest of this section without loss of generality.
 Let us then analyse the dependence of $V$ on the five remaining parameters
 ($\omega$, $S$, $K$, $F$ and $d$).
 \begin{itemize}
 \item
  A simple look at the structure of the integrand in (\ref{vd}),
  or at the final expression (\ref{eqap}),
  suggests that $V\to0$ for $\omega\to\infty$.
  Equation (\ref{static}) displays the
  simplification occurring in the static case $\omega=0$.
  In figures \ref{vwZ} and \ref{vwD}, we plot the ratio
  $V/\mathcal{V}$ between the leading correction
  to the terminal velocity and its bare value, as a function of the frequency.
 \begin{figure}
  \centering
  \includegraphics[height=8cm]{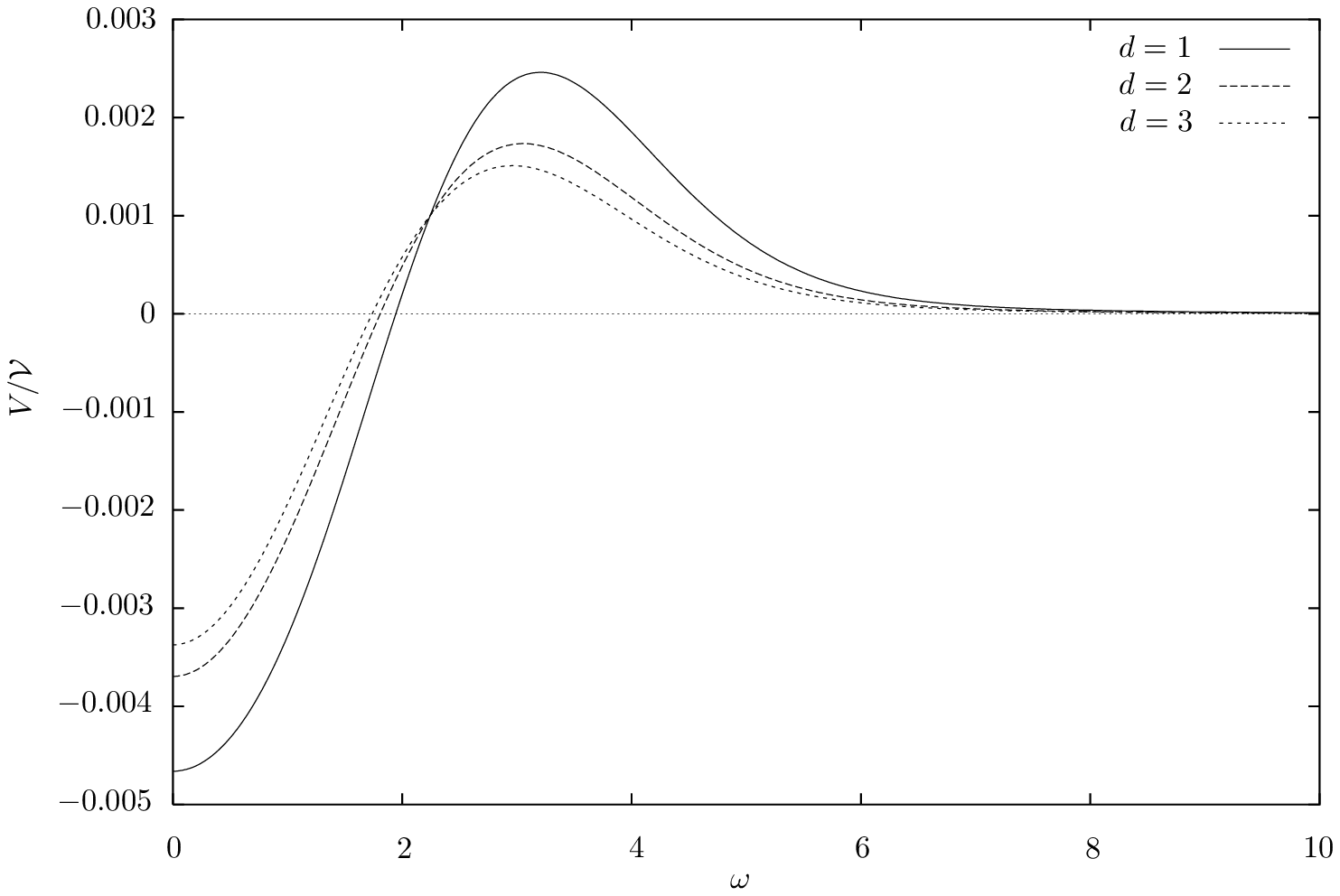}
  \caption{Leading-order correction to the terminal velocity, normalized with the corresponding bare value, as a function of $\omega$, for $d=1$, $2$ and $3$. Here, $\mathcal{P}=1=S$ and $F=0.1=K$.}
  \label{vwZ}
 \end{figure}
 \begin{figure}
  \centering
  \includegraphics[height=8cm]{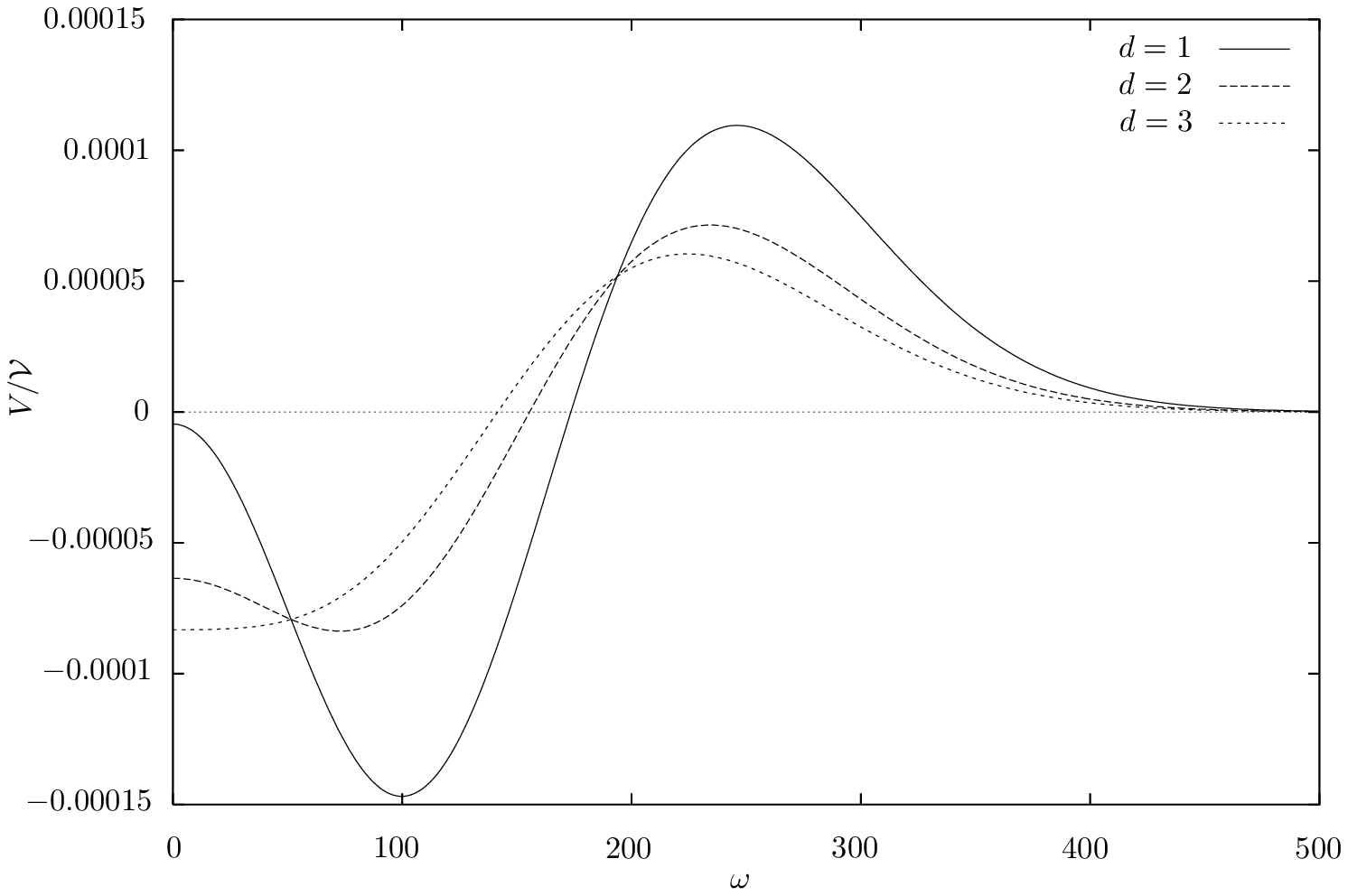}
  \caption{Same as in figure \ref{vwZ} but with $S=10$ and $F=1=K$.}
  \label{vwD}
 \end{figure}
  We notice that small values of the frequency are associated with a negative renormalization, while
  high-frequency flows are characterized by an increased terminal velocity.
  A similar situation can be shown to arise as a result of finite-$T$
  corrections, in the transport of inertial particles by a
  Kraichnan velocity field \cite[][]{OV07}.
  We can thus conclude that (at least for $\mathcal{P}<d^{-1}$),
  if the extension of negative regions in the time correlation profile
  is sufficiently small (large),
  the renormalized streaming is smaller (larger) than the bare one,
  which amounts to saying that the absence (presence) of areas
  of recirculation helps to make the particles travel slower (faster)
  than in the case of still fluid. As we can see from figure \ref{numeric}, this
  conclusion is confirmed also for finite values of the expansion parameter;
  in the case in exam: $SKF^{-2}=4$. Notice that an oscillating exponential
  profile for the time correlation has been utilized in this case, in place
  of the oscillating Gaussian of (\ref{cxt}).
 \begin{figure}
  \centering
  \includegraphics[height=7cm]{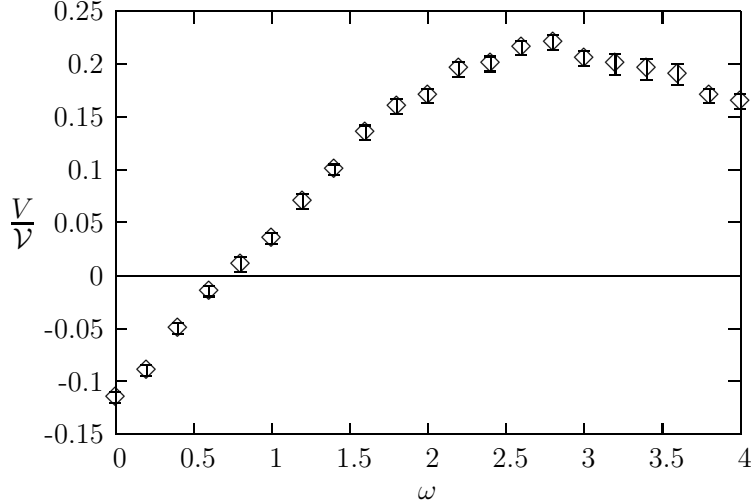}
  \caption{Leading-order correction to the terminal velocity vs.\ $\omega$,
   from a numerical simulation of transport by a 2D fully compressible velocity field.
   Parameters of the simulation: $S=2$, $K=0.5=F$. The size of the spatial domain is $10\,L$.}
  \label{numeric}
 \end{figure}
 \item
  For $T\gg\mathcal{T}_{\mathrm{sw}}$ (see (\ref{T_sweep})) the
  integrand in (\ref{vd}) will depend solely on $\mathcal{T}_{\mathrm{sw}}$.
  Thus, if one studies the zero of expression (\ref{vd}) (i.e.\ the boundary that separates
  accelerating and decelerating situations, at the lowest order), the most
  interesting plot is represented by the separatrix line in the
  plane $\omega$ vs.\ $SKF^{-1}$ for not-too-small values of $K$,
  such that $S$, $K$ and $F$ appear only via this combination.
  In figure \ref{zero}, the lower right area corresponds to
  deceleration and the upper left to acceleration in sweeping.
 \begin{figure}
  \centering
  \includegraphics[height=8cm]{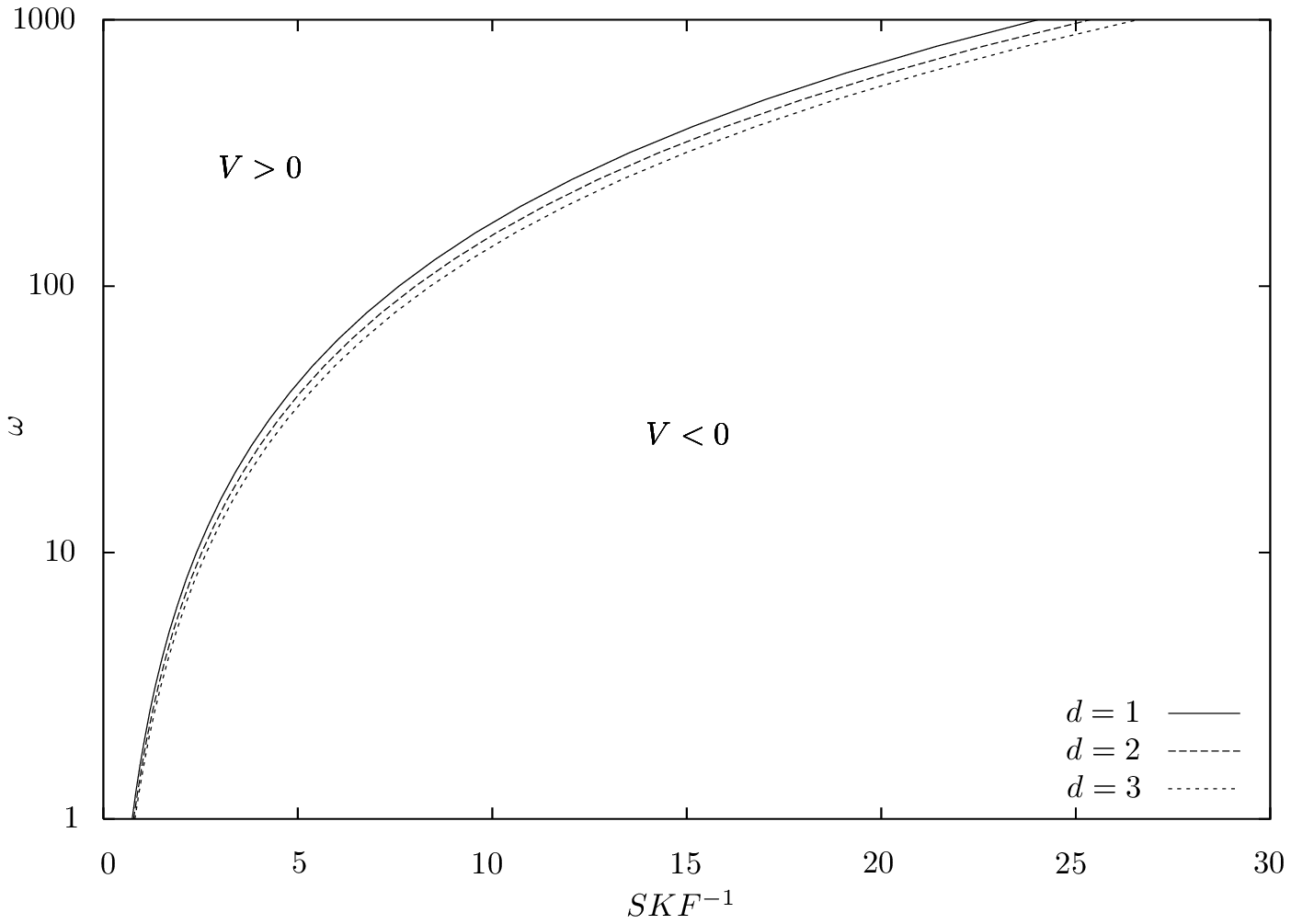}
  \caption{Separatrix between regions of increased (upper left) and decreased (lower right) sweeping in the plane $\omega$ vs.\ $SKF^{-1}$, for $d=1$, $2$ and $3$ (solid, dashed and dotted lines, respectively). The plot namely corresponds to fixed values of $K=1$ and $F=0.1$, but varying these parameters only negligible changes appear (unless $K\ll1$).}
  \label{zero}
 \end{figure}
  Notice that low (respectively, high) values of $SKF^{-1}$
  correspond preferentially to situations of increased
  (respectively, decreased) terminal velocity.
 \item
  The same simplification into a dependence on $SKF^{-1}$ only
  is not valid for the whole expression
  (\ref{vd}) itself, because of the prefactor. Nevertheless, an
  interesting rescaling property can be shown to hold when the
  parameters belong to a specific range. First of all, if one
  considers the ratio $V/\mathcal{V}$, the prefactor $S^3K^3F^{-2}$ in
  (\ref{vd}) becomes $S^2K^2$, i.e.\ $(SKF^{-1})^2F^2$. Now,
  suppose that $T\gg\mathcal{T}_{\mathrm{sw}}$,
  such that, in the exponent of the Gaussian factor,
  $\Delta\gg1$ and thus $\Gamma\simeq(SKF^{-1})^2$.
  Also, suppose that $SKF^{-1}\gg1$, so that
  the same Gaussian is very narrow. This allows us to Taylor expand
  $\psi(\tau)\sim\tau$. Then, starting from a
  situation in which such approximations hold (e.g., the case plotted
  in figure \ref{vwD}), suppose to rescale $SKF^{-1}$ by a factor $\alpha$.
  The expression in square brackets and the simplified Gaussian factor
  remain unchanged upon rescaling $\tau$ by $\alpha^{-2}$,
  and consequently the cosine does not change if $\omega$ is rescaled
  by $\alpha^2$. Keeping into account the $\tau$ factor explicitly
  appearing in (\ref{vd}), the one coming from the Taylor expansion of $\psi(\tau)$, and the integration
  variable differential, one deduces that the integral gets rescaled by
  $(\alpha^{-2})^3=\alpha^{-6}$. By considering also the above-mentioned
  prefactor $(SKF^{-1})^2F^2$, one can argue that, under these approximations: I) upon rescaling
  $SKF^{-1}$ by $\alpha$ while keeping $F$ fixed, figure \ref{vwD}
  simply reproduces itself by rescaling $\omega$ by $\alpha^2$ on the abscissae
  and $V/\mathcal{V}$ by $\alpha^{-4}$ on the ordinates; II) upon rescaling
  $F$ by $\beta$ while keeping $SKF^{-1}$ fixed, figure \ref{vwD}
  simply expands vertically by $\beta^2$ without changing horizontally.\\
  In figures \ref{vme} and \ref{vmd}, $V/\mathcal{V}$ is now plotted
  as a function of $SKF^{-1}$ for different values of the frequency.
 \begin{figure}
  \centering
  \includegraphics[height=8cm]{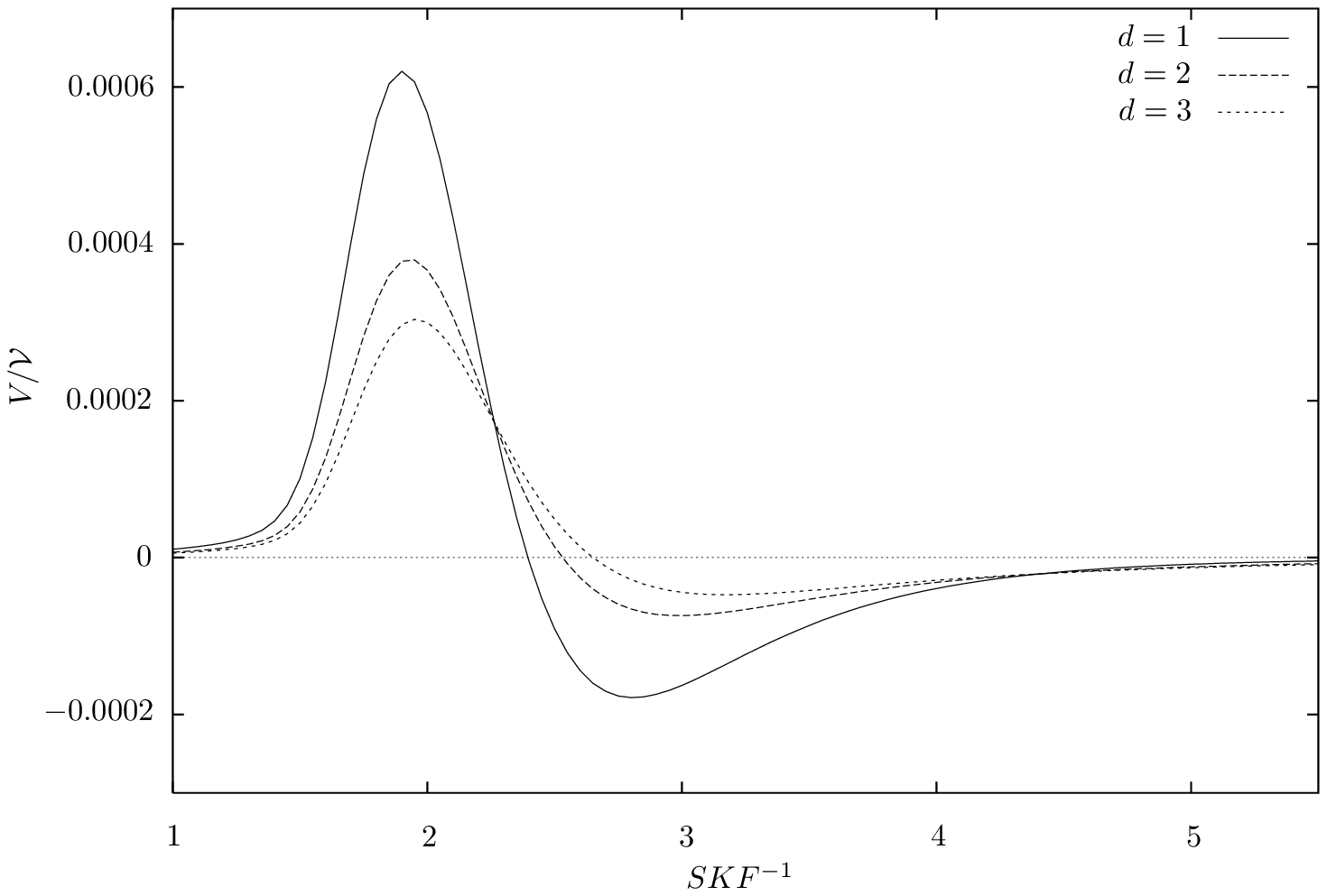}
  \caption{Leading-order correction to the terminal velocity, normalized with the corresponding bare value, as a function of $SKF^{-1}$, for $d=1$, $2$ and $3$. Here, $\mathcal{P}=1$ and $\omega=10$. The plot namely corresponds to fixed values of $K=0.1=F$, and varying these parameters it gets rescaled as explained in the text.}
  \label{vme}
 \end{figure}
 \begin{figure}
  \centering
  \includegraphics[height=8cm]{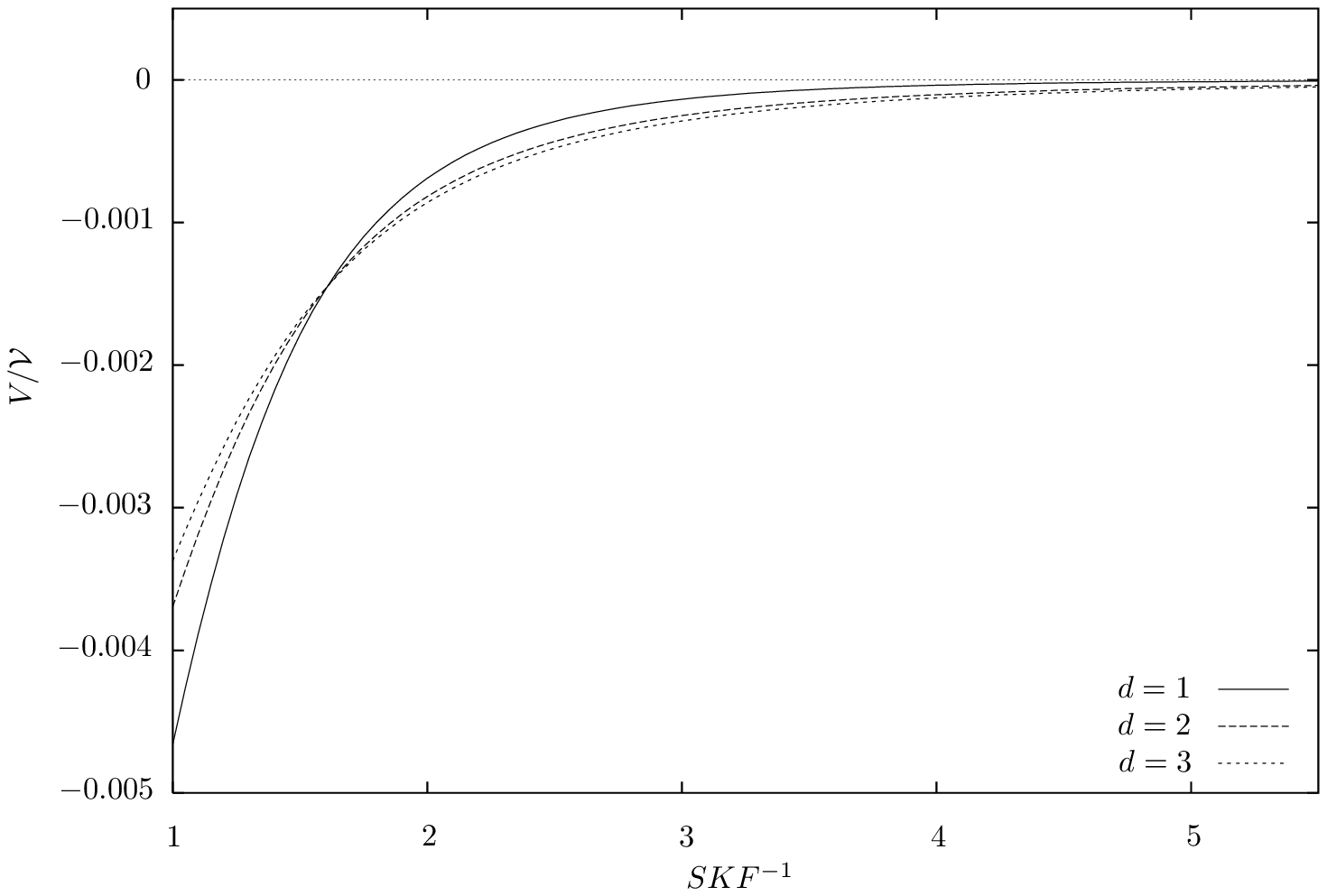}
  \caption{Same as in figure \ref{vme} but for the static case $\omega=0$.}
  \label{vmd}
 \end{figure}
 \item
  The dependence on the dimension seems to be very weak.
  In particular, all our plots show little difference between
  $d=1$, $2$ and $3$, with the one-dimensional case showing
  the largest peaks. Moreover, figures \ref{vwZ}--\ref{vwD} and
  \ref{vme}--\ref{vmd} are peculiar in the sense that the three lines
  always cross in the very same points. We have not been able yet
  to identify the underlying analytical mechanism causing
  this phenomenon, which however seems to be very robust because
  it is also confirmed by plotting e.g.\ a fictitious four-dimensional case.
 \end{itemize}

 \section{Incompressible Gaussian flows} \label{sec:igf}

\begin{figure}
  \centering
  \includegraphics[height=8cm]{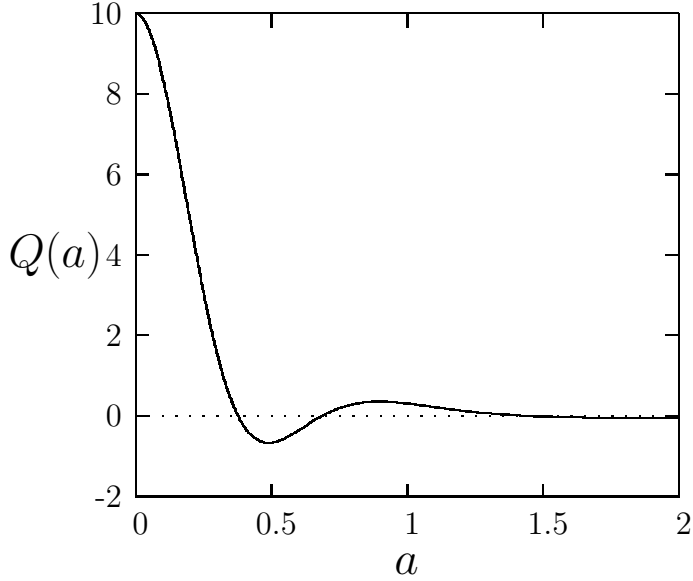}
  \caption{The profile of the function $Q$.}
  \label{maxfig2}
 \end{figure}

 The role of $\omega$, i.e.\ of oscillations in the velocity correlation
 profile, in determining the sign of the terminal velocity renormalization,
 is completely different in the case of a Gaussian incompressible flow. As discussed
 in \cite{M87b}, this requires going to fourth order in the perturbation
 expansion (see (\ref{v3})).\\
 The interest of the problem is rather formal, as non-Gaussianity in realistic
 turbulent flows is likely to play an important role.
 Nevertheless, simple analytical expressions can be obtained in the
 small-sweep regime, with rapid decorrelation imposed through smallness of
 $T$. More precisely, we assume
 \begin{equation} \label{limit}
  SK^2F^{-2}\ll1,\qquad S\gg\max(1,K^2)\;.
 \end{equation}
 In this limit, the decorrelation $\Gamma$ is dominated by the Eulerian correlation
 time $T\ll\tau_{\mathrm{S}}=1$ and the total displacement in a time $T$ will be
 much smaller than the characteristic scale $L=S^{-1}K^{-1}$.\\
 As detailed in appendix \ref{sec:a2}, this allows to expand the integrand in (\ref{v3})
 in both space (the correlations $\langle\mathcal{U}\mathcal{U}\rangle$) and time
 (the propagator $\psi$). Notice that, in these approximations,
 the precise form of the spatial correlation (\ref{rij}) has no influence on
 the sign of the correction, as only
 its second derivative computed at merged points, (\ref{ff}), enters the calculation.
 A simpler expression for the settling velocity renormalization, illustrating the
 changing sign behaviour, is obtained with a temporal correlation in the form:
 \begin{equation} \label{exp}
  h(t)=\ue^{-S|t|}\cos(\omega t)\;.
 \end{equation}
 Substituting into (\ref{semifinal}), we obtain:
 \begin{equation} \label{final}
  \hat{V}\equiv\lim_{t\to\infty}\langle\tilde{v}_d^{\scriptscriptstyle(4)}(t)\rangle=\frac{(d+1)(d+2)\hat{f}^2K^5}{4(d-1)S^2F^2}Q\left(\frac{\omega}{S}\right)\;,
 \end{equation}
 where
 \[Q(a)=\frac{11-94a^2+177a^4-45a^6-7a^8-a^{10}}{(1+a^2)^7}\;.\]
 Using (\ref{eq1.2}), we see that $\hat{V}/\mathcal{V}\propto S^{-3}K^4$, corresponding
 to an $O(S^{-3}K^4)$ renormalization of $\gamma$, to be compared with the $O(S^{-1}K^2)$
 renormalization observed in the fully compressible 1D case \cite[][]{OV07}.
 The plot of the function $Q(a)$ is shown in figure \ref{maxfig2}.
 It was observed in \cite{M87b}, in the case of strong sweep,
 and in the absence of oscillations for $h(t)$, that the correction
 to the settling velocity is positive. We see in figure \ref{maxfig2}
 that the same situation occurs in the small sweep regime, while
 we can also identify a range of values of $\omega$ for which
 the falling velocity is decreased.
 However, the renormalization effect quickly becomes negligible at large $\omega$.
 In any case, we were able to show that, at least in some particular
 instances, the presence or absence of compressibility makes
 the role of $\omega$ the opposite passing from one case to the other.

 \section{Particle diffusivity} \label{sec:pd}

\begin{figure}
  \centering
  \includegraphics[height=8cm]{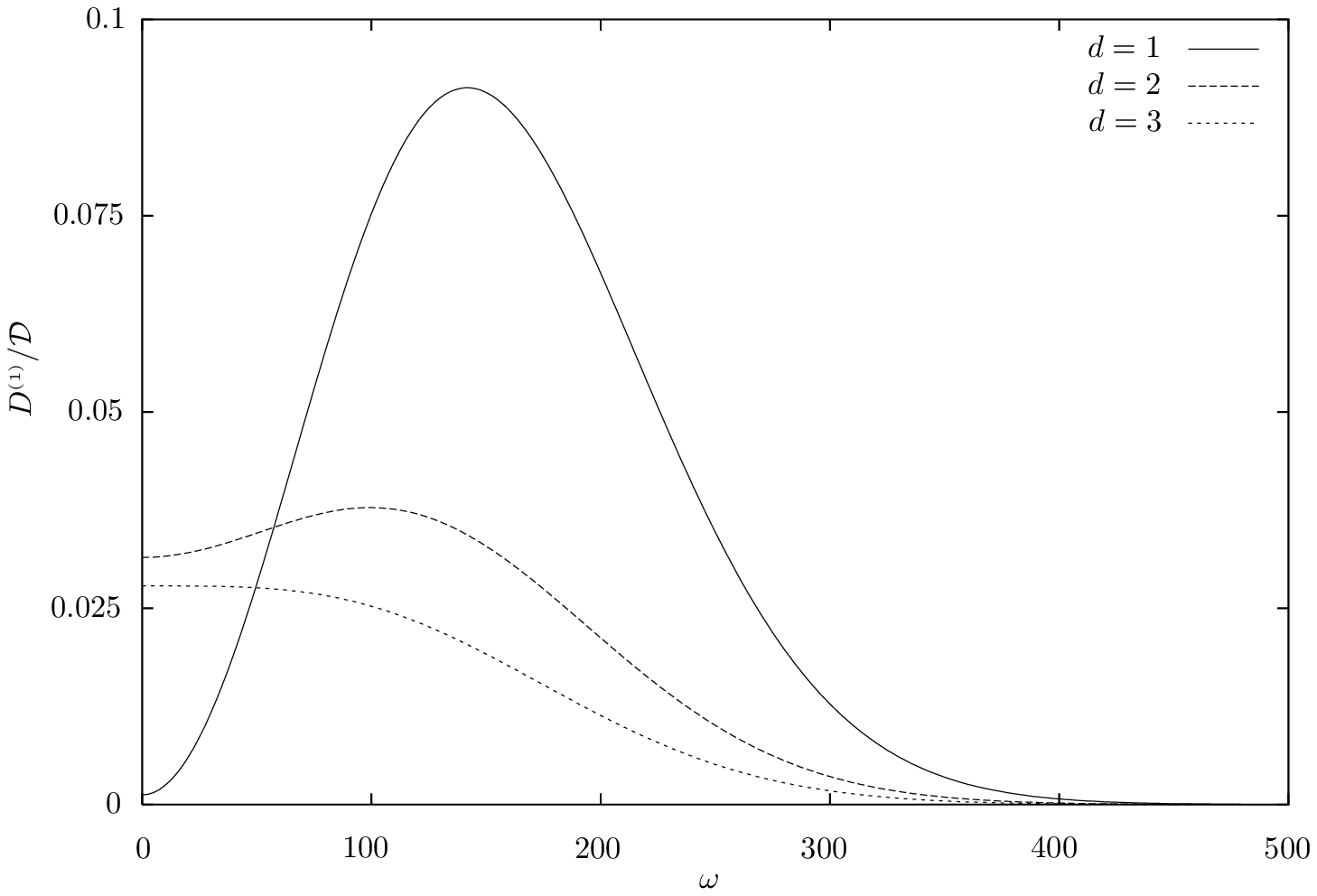}
  \caption{Leading-order expression for the total diffusivity, divided by the corresponding fluid-particle value, as a function of $\omega$, for $d=1$, $2$ and $3$. Here, $S=10$, $\forall K=F$ and $\forall\mathcal{P}$.}
  \label{dwa}
 \end{figure}

\begin{figure}
  \centering
  \includegraphics[height=8cm]{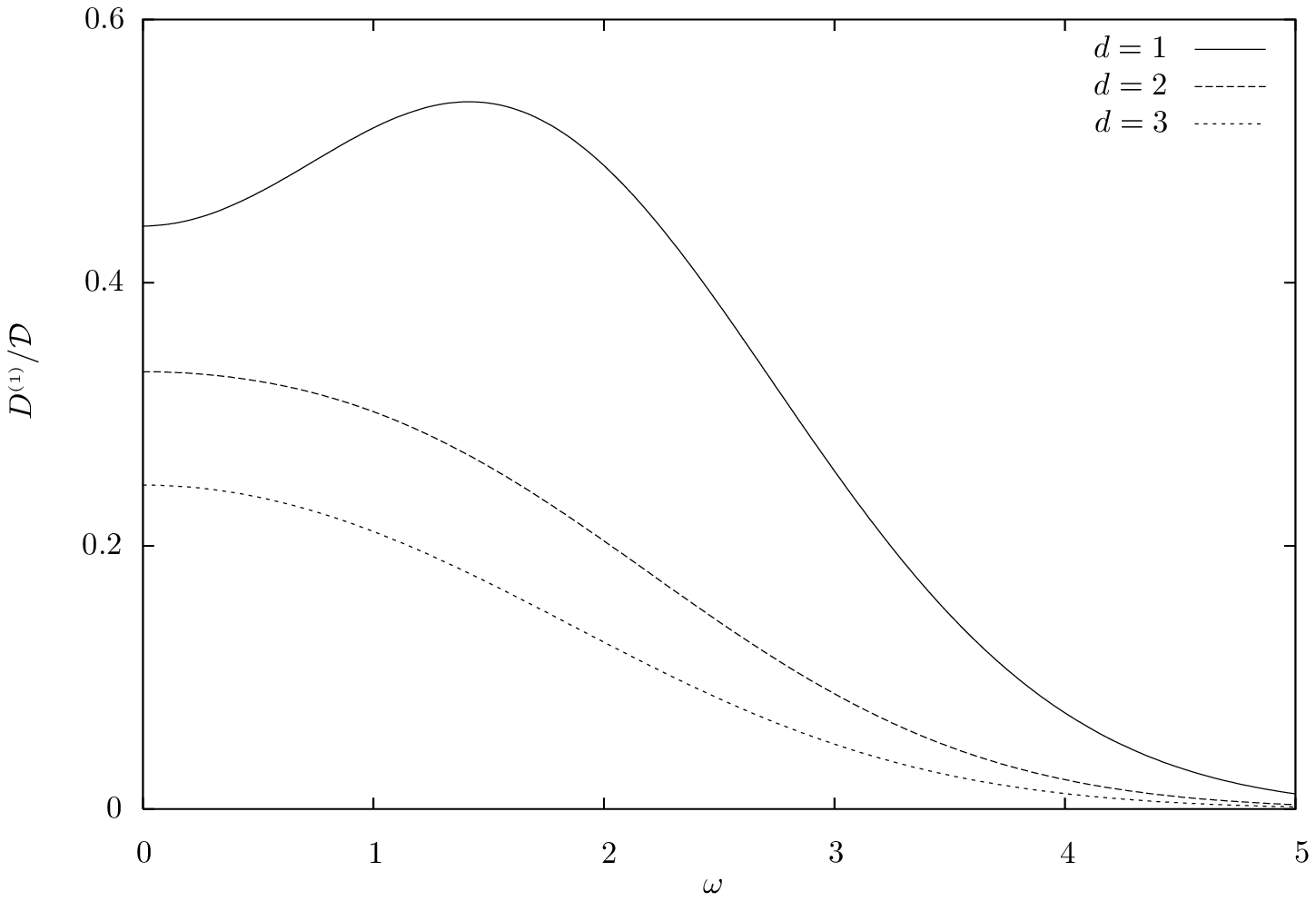}
  \caption{Same as in figure \ref{dwa} but with $S=1$.}
  \label{dwb}
 \end{figure}

\begin{figure}
  \centering
  \includegraphics[height=8cm]{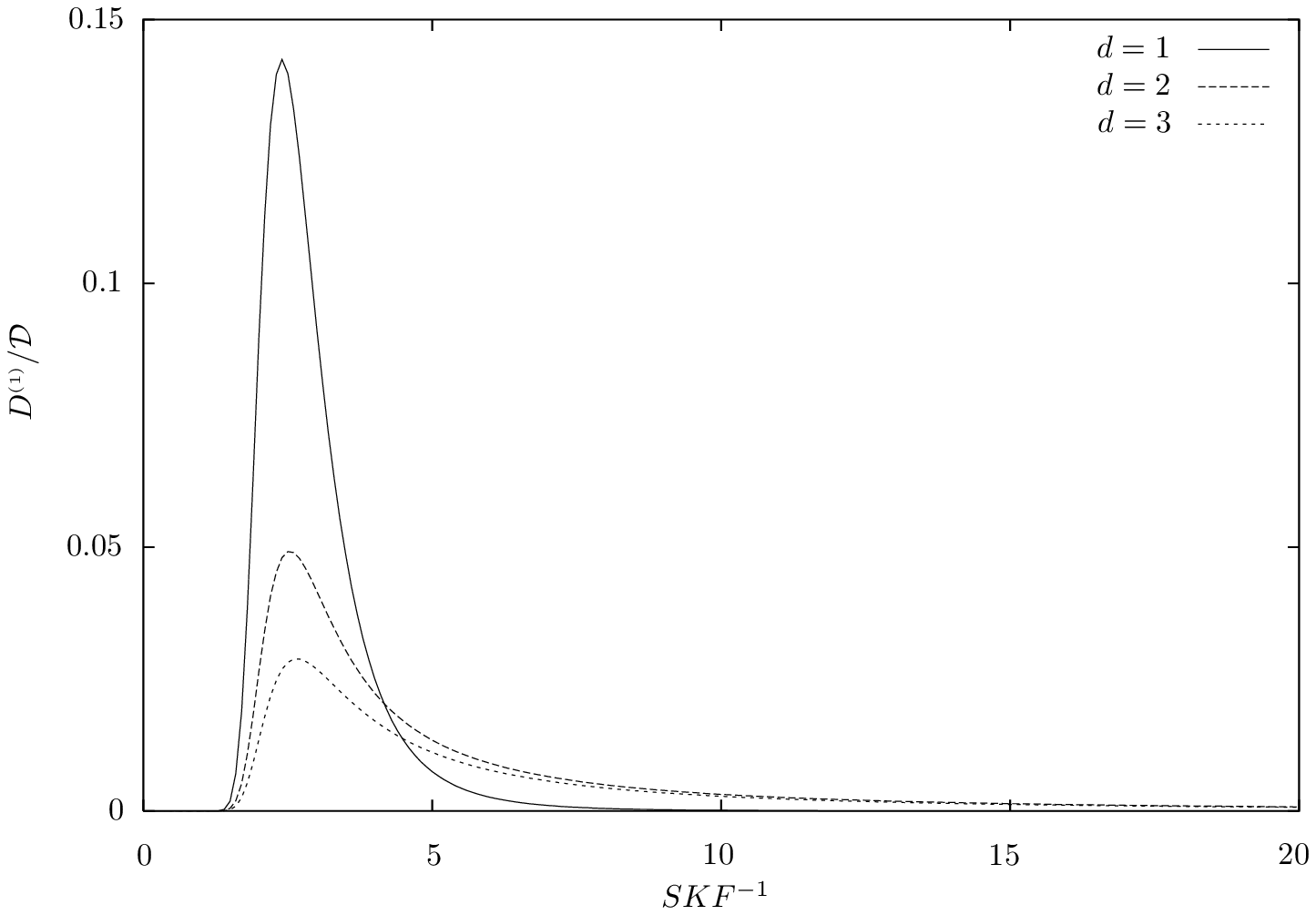}
  \caption{Leading-order expression for the total diffusivity, divided by the corresponding fluid-particle value, as a function of $SKF^{-1}$, for $d=1$, $2$ and $3$. Here, $\omega=10$ and $S=1$, $\forall\mathcal{P}$.}
  \label{dma}
 \end{figure}

\begin{figure}
  \centering
  \includegraphics[height=8cm]{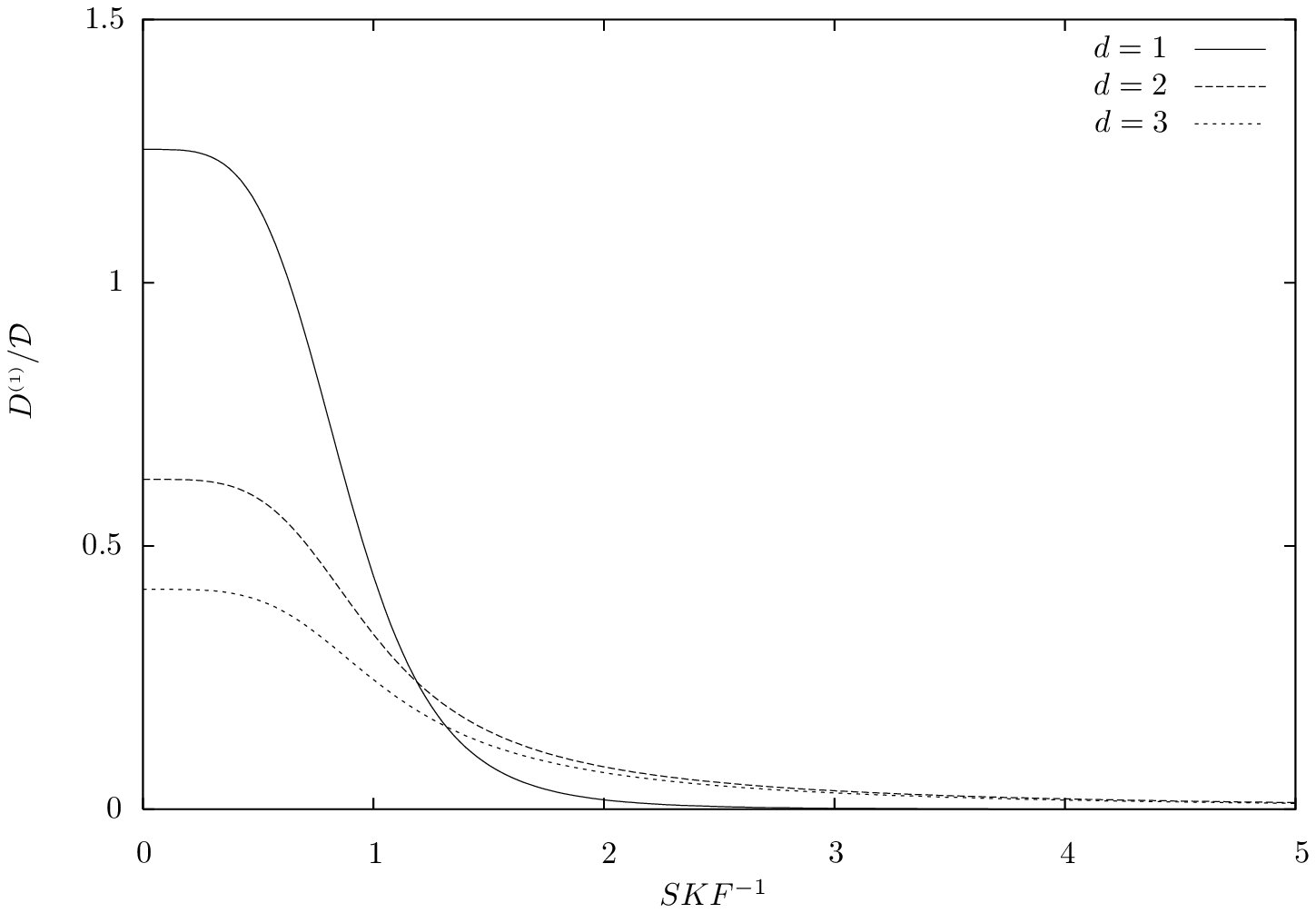}
  \caption{Same as in figure \ref{dma} but for the static case $\omega=0$.}
  \label{dmb}
 \end{figure}

 After computing the terminal velocity, it is interesting to study how the
 particle diffuse. In particular, we would like to find the diffusion coefficient
 \begin{equation} \label{diffus}
  D=\lim_{t\to\infty}\frac{\langle[\bm{x}(t)-\langle\bm{x}\rangle(t)]^2\rangle}{2dt}\;,
 \end{equation}
 to be compared with the diffusivity of a fluid parcel,
 \[\mathcal{D}\sim\sigma_u^2T=S^{-1}.\]
 With this aim, we firstly notice that the term $\bm{x}^{\scriptscriptstyle(0)}$ in (\ref{xbar})
 does not contribute to expression (\ref{diffus}), because it is independent of the
 external random flow. It is thus sufficient to investigate $\tilde{\bm{x}}$
 from (\ref{xtilde}), and namely (using (\ref{mathcal})) its leading order
 \[\tilde{\bm{x}}^{\scriptscriptstyle(1)}(t)=\!\int_0^t\ud t'\,\psi(t-t')\bm{\mathcal{U}}(t')\;.\]
 For zero-mean flows we have $\langle\tilde{\bm{x}}^{\scriptscriptstyle(1)}(t)\rangle=0$, therefore:
 \begin{eqnarray} \label{dtau}
  D^{\scriptscriptstyle(1)}=\lim_{t\to\infty}\frac{1}{2dt}\!\int_0^t\ud t'\,\psi(t-t')\!\int_0^t\ud t''\,\psi(t-t'')\langle\mathcal{U}_i(t')\mathcal{U}_i(t'')\rangle\;.
 \end{eqnarray}
 Looking at (\ref{dtau}), we see that $\psi(t-t')$ and
 $\psi(t-t'')$ differ from $1$ only for $t'$ and $t''$ close to $t$, and
 for $t\to\infty$ we can disregard this contribution.
 Thus, $D^{\scriptscriptstyle(1)}$ is simply the
 diffusivity of a random walker with velocity $\bm{\mathcal{U}}$, and
 we have the standard expression
 \begin{equation} \label{rw}
  D^{\scriptscriptstyle(1)}=\frac{1}{d}\!\int_0^\infty\ud t\,\langle\mathcal{U}_i(0)\mathcal{U}_i(t)\rangle\;.
 \end{equation}
 If we now restore our space-and-time Gaussian oscillating correlation,
 we can replace the average in (\ref{rw}) with the trace of expression (\ref{uijxt}), computed at
 space-time separation $(\bm{\mathcal{V}}t,t)$, and we obtain the result:
 \begin{equation} \label{diff}
  D^{\scriptscriptstyle(1)}=\frac{\sqrt{\pi}S^4}{\sqrt{2}d^2\Gamma^5}\ue^{-\omega^2/2\Gamma^2}\left\{(d-1)\Delta^4+\left[(2d-1)+\left(\frac{\omega}{S}\right)^2\right]\Delta^2+d\right\}\;,
 \end{equation}
 with $\Gamma$ and $\Delta$ given in (\ref{Gamma}).\\
 Notice that the compressibility degree $\mathcal{P}$ disappears in the
 trace $R_{ii}$, and thus does not affect the diffusion coefficient.
 It can easily be seen from (\ref{diff}) that $D^{\scriptscriptstyle(1)}$
 depends on $S$ and $SKF^{-1}$ only, and not on $K$ and $F$ separately.
 Figures \ref{dwa} and \ref{dwb} show the behaviour of the diffusivity
 as a function of the frequency.
 In figures \ref{dma} and \ref{dmb}, the diffusivity is plotted versus $SKF^{-1}$.
 In all cases, plotted is the ratio of the particle-to-fluid diffusivities, $D^{\scriptscriptstyle(1)}/\mathcal{D}$.

 As a further step, one can also investigate the different diffusive behaviour
 between the ``sweeping'' direction and the orthogonal one(s) (if any).
 In particular, one obtains
 \[D^{\scriptscriptstyle(1)}_{\parallel}\equiv\lim_{t\to\infty}\frac{\langle[\tilde{x}_d^{\scriptscriptstyle(1)}(t)]^2\rangle}{2dt}=\frac{\sqrt{\pi}S^4}{\sqrt{2}d^2\Gamma^5}\ue^{-\omega^2/2\Gamma^2}\left\{(1-\mathcal{P})\Delta^4+\left[(2-\mathcal{P})+\mathcal{P}\left(\frac{\omega}{S}\right)^2\right]\Delta^2+1\right\}\]
 and
 \begin{eqnarray*}
  D^{\scriptscriptstyle(1)}_{\perp}\equiv D^{\scriptscriptstyle(1)}-D_{\parallel}^{\scriptscriptstyle(1)}&=&\frac{\sqrt{\pi}S^4}{\sqrt{2}d^2\Gamma^5}\ue^{-\omega^2/2\Gamma^2}\bigg\{(d+\mathcal{P}-2)\Delta^4\\
  &&\hspace{2.7cm}\left.+\left[(2d+\mathcal{P}-3)+(1-\mathcal{P})\left(\frac{\omega}{S}\right)^2\right]\Delta^2+(d-1)\right\}
 \end{eqnarray*}
 (which correctly vanishes for $d=1$ because there $\mathcal{P}=1$).\\
 It is worth noticing that the compressibility degree now appears again
 in the two previous expressions: only their sum is independent of it.

 \subsection{Comparison between drift and diffusion displacements}

\begin{figure}
  \centering
  \includegraphics[height=8cm]{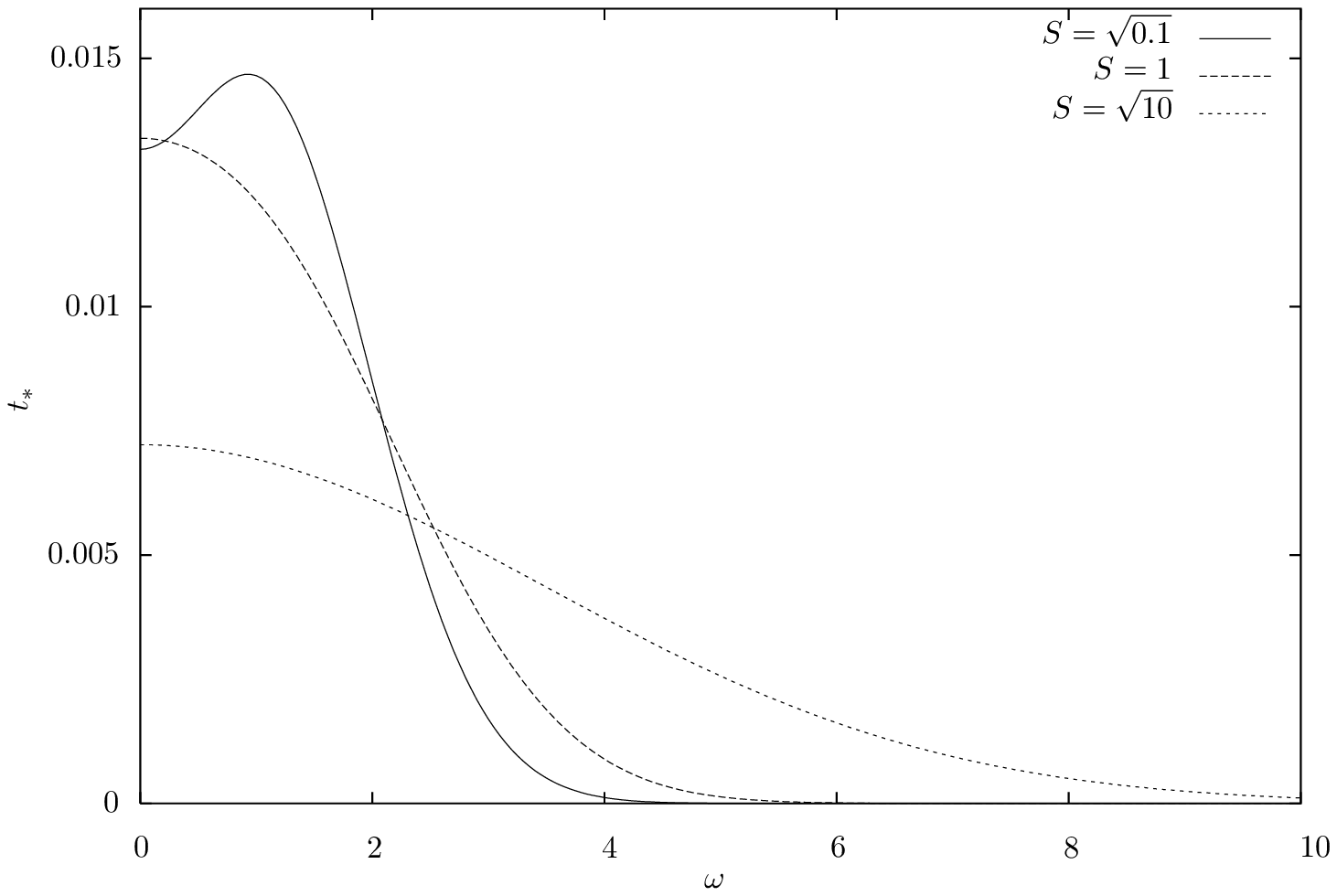}
  \caption{Crossover time, between diffusion- and drift-dominated particle displacement, as a function of $\omega$. Here, $d=2$, $\mathcal{P}=1$, $F=0.1$ and $SKF^{-1}=1$.}
  \label{fig:tc}
 \end{figure}

 It is now possible to compare the contributions to a particle
 displacement given by drift and by diffusion. At the time $t$
 large enough to forget about initial conditions, the typical
 displacement due to drift is about $(\mathcal{V}+V)t$, while the
 diffusive one is of the order of
 $\sqrt{2dD^{\scriptscriptstyle(1)}t}$. The former clearly
 dominates at large times, while the latter represents the leading
 value of the displacement at short times. We can thus define the
 crossover time
 \[t_*\equiv\frac{2dD^{\scriptscriptstyle(1)}}{(\mathcal{V}+V)^2}\;,\]
 after which drift represents the dominant mechanism and diffusion
 can be neglected.\\
 Focusing on the two-dimensional potential-flow case,
 in figure \ref{fig:tc} such crossover time (remember that, in our convention,
 it is adimensionalized with $\gamma$) is plotted as a function of
 the frequency for three different values of the Stokes number.
 Notice the peak that can occur in $t_*$ at a critical value of
 $\omega$ when the other parameters assume certain values.

 \section{Conclusions and perspectives} \label{sec:c}

 We investigated the drift and diffusion behaviour of inertial particles in
 random flows, in the limit in which the
 expansion ``\`a la Maxey'' works, i.e.\ in the presence of strong
 sweeping, or of an underlying flow varying rapidly,
 such that the resulting motion at the scales of interest
 is almost rectilinear.  This picture can be
 applied both to the settling of heavy particles due to gravity,
 and, more interestingly, to the motion of floaters, e.g.\ on a
 water surface, under the action of a streaming wind.

 Imposing an oscillating Gaussian form to the external flow, we
 found the analytical expressions of the renormalized terminal
 velocity and of the effective diffusivity, as functions of the
 oscillation frequency and of the Stokes, Kubo and Froude numbers.
 The latter three quantities often appear not independently from
 one another, but rather as an overall combination. For the
 terminal velocity, we found that, if the flow is compressible, small values of the frequency are
 associated with a negative renormalization with respect to the
 corresponding value in still fluids, while high-frequency flows
 are characterized by an increased asymptotic speed. In some cases,
 the opposite behaviour holds as a function of the parameter $SKF^{-1}$,
 or as a function of $\omega$ itself if the flow is Gaussian and incompressible.
 The effective diffusivity can show some ``resonant'' peaks as a
 function both of $\omega$ and of $SKF^{-1}$. The role played by
 the compressibility degree $\mathcal{P}$ is apparently quite simple: it
 appears linearly in the leading correction to the renormalized terminal velocity (so that,
 for incompressible flows, higher-order effects have to considered),
 and it does not influence the total effective diffusivity,
 because the dependence on $\mathcal{P}$ shown by the components
 of the diffusivity parallel and orthogonal to the sweeping direction
 is such as to create an overall compensation. However, we were
 able to show that the value of $\mathcal{P}$ can have relevant
 consequences on the role played by $\omega$ on the renormalized streaming.
 In particular, we can conclude that, for compressible flows (at least when
 the compressibility degree is not too high, but neither too low in order
 to have non-negligible effects), the introduction of time oscillations
 in the velocity correlation function mimics the presence of areas of
 recirculation, or in other words of negatively-correlated regions in the flow,
 and is associated with an increase in the particle terminal velocity;
 on the contrary, a decrease is found for slow or no oscillations
 (for which the Eulerian time correlation is positive). In all cases,
 we also found that the dependence on the dimension ($d=1,2,3$) is quite weak.

 \begin{acknowledgements}
  MMA is grateful to Keck Foundation for financial support and to
  Grisha Falkovich for useful discussions.
 \end{acknowledgements}

 \appendix

 \section{Exact results for the compressible case} \label{sec:a1}

 The result of the integral in (\ref{vd}), computed with MATHEMATICA (which makes use of complex variables), is:
 \begin{eqnarray} \label{eqap}
  V&=&\frac{\mathcal{P}S^5K^3}{dF^2\Gamma^7}\Bigg\{-\sqrt{\frac{\pi}{2}}S^2\,\Re\Bigg[\Bigg((d-1)\Delta^4+\bigg((2d+1)+\Big(\frac{\omega+\ui}{S}\Big)^2\bigg)\Delta^2+(d+2)\Bigg)\times\nonumber\\
  &&\hspace{0.5cm}\times(1-\ui\omega)\ue^{(1-\ui\omega)^2/2\Gamma^2}\Bigg(1-\mathrm{erf}\left(\frac{1-\ui\omega}{\sqrt{2}\Gamma}\right)\Bigg)\Bigg]-\Gamma\Delta^2+\sqrt{\frac{\pi}{2}}S^2\omega\ue^{-\omega^2/2\Gamma^2}\times\nonumber\\
  &&\hspace{0.5cm}\times\Bigg[(d-1)\Delta^4+\Bigg((2d+1)+\left(\frac{\omega}{S}\right)^2\Bigg)\Delta^2+(d+2)\Bigg]\left|\mathrm{erf}\left(\frac{\ui\omega}{\sqrt{2}\Gamma}\right)\right|\Bigg\}\;,
 \end{eqnarray}
 where $\mathrm{erf}$ is the usual error function, $\Re$ denotes the real part, and $\Gamma$, $\Delta$ were defined in (\ref{Gamma}).\\
 In the static case ($\omega=0$), this expression simplifies to:
 \begin{eqnarray} \label{static}
  V|_{\omega=0}&=&-\frac{\mathcal{P}S^5K^3}{dF^2\Gamma^7}\left\{\Gamma\Delta^2+\sqrt{\frac{\pi}{2}}S^2\ue^{1/2\Gamma^2}\left[1-\mathrm{erf}\left(\frac{1}{\sqrt{2}\Gamma}\right)\right]\times\right.\\
  &&\hspace{2.5cm}\left.\times\left[(d-1)\Delta^4+\left((2d+1)-\frac{1}{S^2}\right)\Delta^2+(d+2)\right]\right\}\;.\nonumber
 \end{eqnarray}

\section{Technical details for the incompressible case} \label{sec:a2}

\begin{figure}
  \centering
  \includegraphics[width=8cm]{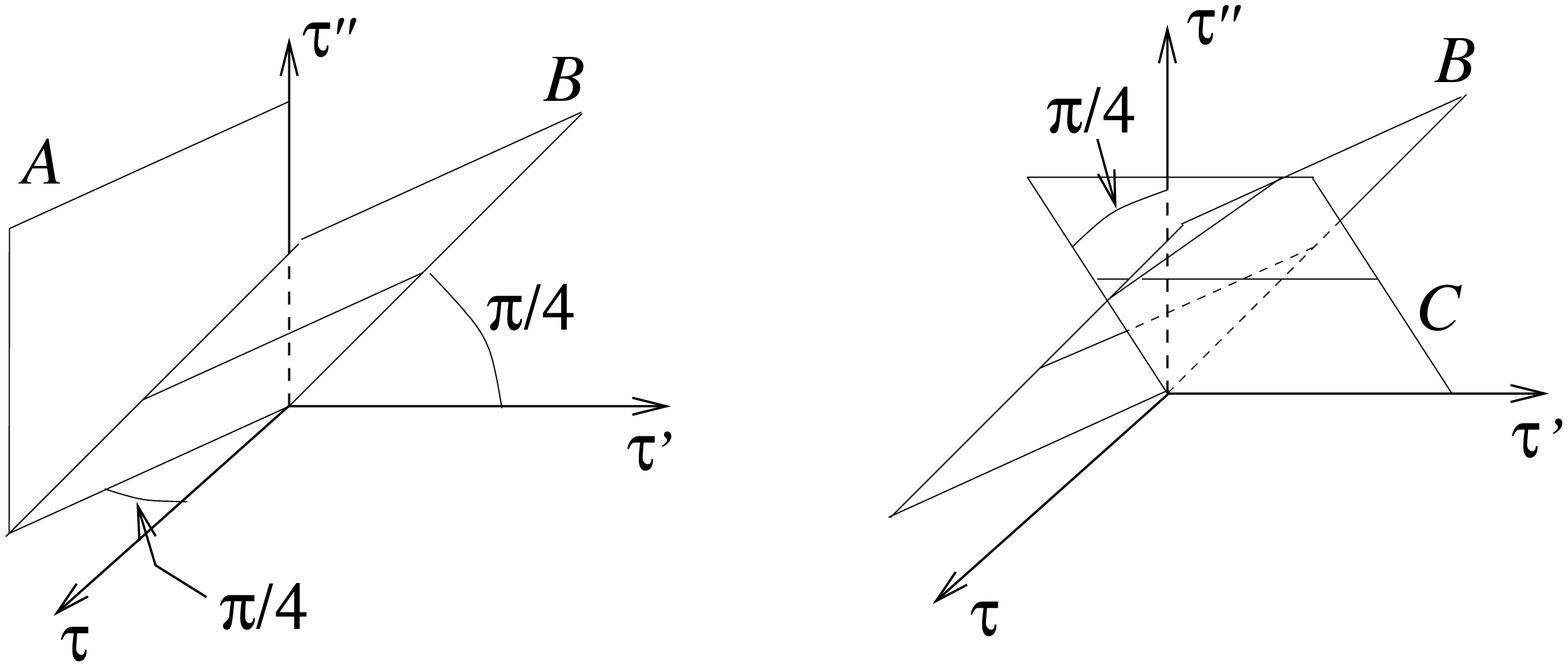}
  \caption{The domain for (\ref{int1}) is the wedge between planes $A$ and $B$ and the coordinate plane $\tau'\tau''$.
   The domain for (\ref{int2}) is the wedge between planes $B$ and $C$ and again the coordinate plane $\tau'\tau''$.
   The intersection of both planes $B$ and $C$ with the coordinate plane $\tau\tau''$ is the line $\tau''=\tau$.}
  \label{maxfig1}
 \end{figure}

 Setting $\mathcal{P}=0$, we can rephrase (\ref{uij}) following \cite{M87b} and write
 \begin{equation} \label{rij}
  R_{ij}(\bm{x},t)=\left\{g(x)\delta_{ij}+[f(x)-g(x)]\frac{x_ix_j}{x^2}\right\}h(t)\;,
 \end{equation}
 where $f(x)$ and $g(x)$ are such as to ensure incompressibility, and we assume $h(0)=1$.
 In this way, the integrand of (\ref{v3}), with $i=d$, can be rewritten as:
 \begin{eqnarray} \label{fg}
  H(t''',t',t'',t)&\equiv&\partial_j\partial_l\langle\mathcal{U}_d\mathcal{U}''_k\rangle\partial'_k\langle\mathcal{U}'_j\mathcal{U}'''_l\rangle\\
  &=&\mathrm{sgn}(t'-t''')h(t-t'')h(t'-t''')\left[\frac{d+1}{r}f'(r)g'(s)+\frac{d+1}{d-1}f''(r)f'(s)\right]\;,\nonumber
 \end{eqnarray}
 with $s=\mathcal{V}|t'-t'''|$ and $r=\mathcal{V}|t-t''|$
 (primes on $f$ and $g$ now indicate differentiation with respect to the corresponding variables).
 Notice that $H$ is invariant under changes of third and fourth variables,
 while it flips its sign when changing first and second variables;
 homogeneity also implies that $H(t''',t',t'',t)$ is a function of $t-t''$ and $t'-t'''$ only.\\
 The first condition assumed in (\ref{limit}), corresponding to a quasi-uniform field,
 allows us to expand the functions $f$ and $g$ in (\ref{rij}) in powers of the spatial variable:
 \[R_{ij}(\bm{x},t)\simeq\left[\frac{1}{d}+\frac{1}{2}g''(0)x^2\right]\delta_{ij}+\frac{1}{2}[f''(0)-g''(0)]x_ix_j\]
 (here, $f'(0)=0=g'(0)$ to ensure smoothness, and $f(0)=g(0)=d^{-1}$ by construction).
 The incompressibility constraint gives us $g''(0)=[(d+1)/(d-1)]f''(0)$.
 Reminding the meaning of $L$ and our adimensionalization of lengths in (\ref{eq1.2}), we can introduce
 \begin{equation} \label{ff}
  \hat{f}\equiv L^2f''(0)\ \Longleftrightarrow\ f''(0)=S^2K^2\hat{f}\;,
 \end{equation}
 where $\hat{f}$ is now $O(1)$.\\
 Looking at (\ref{fg}), we see that we must expand the first-order derivatives; this gives us:
 \begin{eqnarray} \label{h}
  H(t''',t',t'',t)&=&(t'-t''')\mathcal{V}h(t-t'')h(t'-t''')\left[(d+1)f''(0)g''(0)+\frac{d+1}{d-1}f''(0)^2\right]\nonumber\\
  &=&\frac{(d+1)(d+2)S^5K^5\hat{f}^2}{(d-1)F^2}h(t-t'')h(t'-t''')(t'-t''');.
 \end{eqnarray}
 We then see that the contribution from the space correlation has always the same sign.\\
 The integrals in (\ref{v3}) can be written in the equivalent form:
 \begin{eqnarray*}
  \langle\tilde{v}_i^{\scriptscriptstyle(4)}(t)\rangle&=&\!\int_0^t\ud\tau\,\psi(\tau)\left\{\!\int_{-\tau}^{t-\tau}\ud\tau'\int_{\tau+\tau'}^t\ud\tau''\,\psi(\tau+\tau')\psi(\tau''-\tau-\tau')\right.\\
  &&\left.+\!\int_{\tau}^t\ud\tau''\int_{\tau''-\tau}^{t-\tau}\ud\tau'\,\psi(\tau''-\tau)\psi(\tau+\tau'-\tau'')\right\}H(\tau',0,0,\tau'')\;,
 \end{eqnarray*}
 leading to the expression for
 $\hat{V}\equiv\lim_{t\to\infty}\langle\tilde{v}_d^{\scriptscriptstyle(4)}(t)\rangle$:
 \begin{eqnarray*}
  \hat{V}&=&\!\int_0^{\infty}\ud\tau\,\psi(\tau)\left\{\!\int_{-\tau}^{\infty}\ud\tau'\int_{\tau+\tau'}^{\infty}\ud\tau''\,\psi(\tau+\tau')\psi(\tau''-\tau-\tau')\right.\\
  &&\left.+\!\int_{\tau}^{\infty}\ud\tau''\int_{\tau''-\tau}^{\infty}\ud\tau'\,\psi(\tau''-\tau)\psi(\tau+\tau'-\tau'')\right\}H(\tau',0,0,\tau'')\;.
 \end{eqnarray*}
 We can eliminate the integral in $\tau$ rewriting (see figure \ref{maxfig1}):
 \begin{eqnarray} \label{int1}
  \int_0^\infty\ud\tau\int_{-\tau}^{\infty}\ud\tau'\int_{\tau+\tau'}^{\infty}\ud\tau''=\int_0^{\infty}\ud\tau''\int_{-\infty}^{\tau''}\ud\tau'\int_{\max(0,-\tau')}^{\tau''-\tau'}\ud\tau
 \end{eqnarray}
 and
 \begin{eqnarray} \label{int2}
  \int_0^{\infty}\ud\tau\int_{\tau}^{\infty}\ud\tau''\int_{\tau''-\tau}^{\infty}\ud\tau'=\!\int_0^{\infty}\ud\tau''\int_0^{\infty}\ud\tau'\int_{\max(0,\tau''-\tau')}^{\tau''}\ud\tau\;.
 \end{eqnarray}
 Carrying on the integrals in $\tau$ by exploiting (\ref{h}):
 \begin{eqnarray} \label{semifinal}
  \hat{V}&=&-\frac{(d+1)(d+2)S^5K^5\hat{f}^2}{(d-1)F^2}\!\int_0^\infty\ud\tau''\,h(\tau'')\bigg\{\!\int_{\-\infty}^0\ud\tau'\,G_1(\tau',\tau'')\nonumber\\
  &&+\!\int_0^{\tau''}\ud\tau'\,[G_2(\tau',\tau'')+G_4(\tau',\tau'')]+\!\int_{\tau''}^\infty\ud\tau'\,G_3(\tau',\tau'')\bigg\}\tau'h(-\tau')\;,
 \end{eqnarray}
 where the kernels $G_{\mu}$ ($\mu=1,\ldots,4$) get the following, simple expressions as
 the second condition in (\ref{limit}) allows us to Taylor expand the propagators $\psi$ in $\tau'$ and $\tau''$:
 \begin{eqnarray*}
  G_1(\tau',\tau'')&\equiv&\!\int_{-\tau'}^{\tau''-\tau'}\ud\tau\,\psi(\tau)\psi(\tau+\tau')\psi(\tau''-\tau-\tau')\simeq-\frac{\tau'{\tau''}^3}{6}+\frac{{\tau''}^4}{12}\;,\\
  G_2(\tau',\tau'')&\equiv&\!\int_0^{\tau''-\tau'}\ud\tau\,\psi(\tau)\psi(\tau+\tau')\psi(\tau''-\tau-\tau')\simeq-\frac{{\tau'}^4}{12}+\frac{{\tau'}^3\tau''}{6}-\frac{\tau'{\tau''}^3}{6}+\frac{{\tau''}^4}{12}\;,\\
  G_3(\tau',\tau'')&\equiv&\!\int_0^{\tau''}\ud\tau\,\psi(\tau)\psi(\tau''-\tau)\psi(\tau-\tau''+\tau')\simeq\frac{\tau'{\tau''}^3}{6}-\frac{{\tau''}^4}{12}\;,\\
  G_4(\tau',\tau'')&\equiv&\!\int_{\tau''-\tau'}^{\tau''}\ud\tau\,\psi(\tau)\psi(\tau''-\tau)\psi(\tau-\tau''+\tau')\simeq-\frac{{\tau'}^4}{12}+\frac{{\tau'}^3\tau''}{6}\;.
 \end{eqnarray*}
 The desired result can then be obtained from (\ref{semifinal})
 by specifying the exact form of the temporal correlation $h$;
 imposing (\ref{exp}), we finally get (\ref{final}).

\end{document}